\documentstyle[preprint,aps,prl]{revtex}

\begin{document}
\draft
\preprint{}
\title{
Gauge-Invariant Formulation of Fermi's Golden Rule: \\
Application to High-Field Transport in Semiconductors
}
\author{Emanuele Ciancio and Fausto Rossi}
\address{
Istituto Nazionale per la Fisica della Materia (INFM) and \\ 
Dipartimento di Fisica, Politecnico di Torino \\
Corso Duca degli Abruzzi 24, 10129 Torino, Italy \\[1ex]
FRossi@Athena.PoliTo.It
}

\date{\today}
\maketitle

\begin{abstract}

A gauge-invariant formulation of Fermi's Golden rule is proposed. We shall
rivisit the conventional description of carrier-phonon scattering in the
presence of high electric fields by means of a gauge-invariant
density-matrix approach. We show that the so-called {\it intracollisional
field effect ---as usually accounted for--- does not exist}: it is simply an
artifact due to the neglect of the time variation of the basis states
which, in turn, leads to a ill-defined Markov limit in the carrier-phonon
interaction process. This may account for the surprisingly good agreement
between semiclassical and rigorous quantum-transport calculations.

\end{abstract}


\narrowtext

Since the early days of quantum mechanics \cite{Bloch} the field-induced 
coherent dynamics of an electron wavepacket within a crystal, known as 
{\it Bloch oscillations} (BO),
has attracted significant and increasing interest \cite{BO}. 
Indeed, the problem of properly describing the scattering-free motion of an
electron in a solid has led to a three-decade controversy on the existence 
of BO \cite{Nenciu}; 
This originated from the different approaches employed for the description 
of the applied field, namely {\it the vector-potential or 
accelerated-Bloch-state picture} \cite{Houston} and {\it the scalar-potential 
or Wannier-Stark description} \cite{Wannier}.
As discussed in \cite{SL}, these two pictures are now recognized to be 
fully equivalent, since 
they correspond to different quantum-mechanical representations
connected by a gauge transformation.

The presence of scattering as well as tunneling processes strongly modifies
such ideal BO scenario \cite{scat-tunn}. In particular, non-elastic 
interaction mechanisms ---like carrier-LO phonon scattering--- tend to 
spoil such coherent dynamics, leading to a nearly semiclassical or 
Boltzmann-like transport picture.
In the presence of strong electric fields, however, the use of the 
conventional scattering picture ---involving transitions between field-free
Bloch states within Fermi's golden rule--- becomes questionable.

As originally pointed out by Levinson \cite{Levinson} and
by Barker and Ferry \cite{BF}, the effect of the field during the scattering 
process, usually referred to as {\it intracollisional field effect} (ICFE),
may lead to significant deviations from the semiclassical scenario.
On the one hand, 
the role played by the ICFE has been extensively investigated by means of 
rigorous quantum-transport approaches 
\cite{Brunetti,Bertoncini,Herbst,Hader}. 
Their application, however, was often limited to highly simplified physical 
models and conditions, thus preventing from any quantitative comparison with 
experiments. 
On the other hand, strong effort has been devoted to incorporate the 
ICFE within 
conventional ---and more realistic--- Monte Carlo simulations 
\cite{MC}. 
In this case, the basic idea is that,
due to the field-induced carrier drift, energy conservation in the 
scattering process is relaxed; as a consequence, the delta function of the 
Fermi's golden rule is replaced by broad spectral functions \cite{GMC}. 
We stress that this scenario, intimately related to the vector-potential 
or accelerated picture, has no counterpart in the scalar-potential one. 
Indeed, within the
Wannier-Stark basis there is no carrier drift, and energy conservation is 
preserved.
It is thus clear that such an effective semiclassical description of 
the ICFE is 
not gauge invariant \cite{gauge-invariance}.

Aim of this paper is to explain and remove this apparent contraddiction 
by providing a rigorous ---i.e., gauge-invariant--- formulation of Fermi's 
golden rule. 
Our analysis will show that 
the ICFE, as usually accounted for within semiclassical Monte Carlo 
simulations \cite{GMC}, does not exist: it is simply an
artifact due to the neglect of the time variation of the basis states
which, in turn, leads to a ill-defined Markov limit in the carrier-phonon
interaction process. This may account for the surprisingly good agreement
between semiclassical and rigorous quantum-transport calculations 
\cite{Brunetti,Herbst}.

In order to properly describe high-field transport in semiconductors, 
an electron-phonon system can be considered, whose 
Hamiltonian can be schematically written as 
\begin{equation}\label{H}
{\bf H} = {\bf H}_\circ + {\bf H}' = \left({\bf H}_c + {\bf H}_p\right) + 
{\bf H}_{cp}.
\end{equation}
Here, the single-particle term ${\bf H}_\circ$ includes the free-carrier 
and phonon Hamiltonians while the 
many-body contribution ${\bf H}'$ accounts for carrier-phonon coupling.
More specifically, the free-carrier Hamiltonian
\begin{equation}\label{H_c}
{\bf H}_c = 
{\left(-i\hbar\nabla_{\bf r} -{e\over c} {\bf A}({\bf r},t;\eta)\right)^2
\over 2 m_\circ} + e\varphi({\bf r},t;\eta) + V^l({\bf r})
\end{equation}
describes the non-interacting carrier system within the periodic crystal 
potential $V^l({\bf r})$ in the presence of a homogeneous external field 
${\bf F}$,
the latter being included in a fully gauge-invariant form through the 
following vector and scalar potentials:
\begin{equation}\label{vecscalpot}
{\bf A}({\bf r},t;\eta) = -c\eta {\bf F}\,t \ , \qquad 
\varphi({\bf r};\eta) = (\eta-1){\bf F} \cdot {\bf r} \ .
\end{equation}
Here, the gauge freedom is expressed in terms of the transformation 
parameter $\eta$. 
Indeed, it is easy to verify that for any value of $\eta$ 
---corresponding to a different Hamiltonian ${\bf H}_c$ in Eq.~(\ref{H_c})--- 
we are describing exactly the same electric field ${\bf F}$.
In particular, for $\eta = 0$ and $\eta = 1$ one recovers the conventional 
scalar- and vector-potential formulations, respectively, which in turn 
correspond to the well-known Wannier-Stark and Bloch-oscillation pictures 
\cite{SL}.

Let us finally introduce the carrier-phonon interaction Hamiltonian:
\begin{equation}\label{H_cp}
{\bf H}_{cp} = \sum_{\bf q} \gamma_{\bf q} \left[
b^{ }_{\bf q} e^{i{\bf q\cdot r}} +
b^{\dagger}_{\bf q} e^{-i{\bf q\cdot r}} 
\right],
\end{equation}
where the operators $b^{\dagger}_{\bf q}$ ($b_{\bf q}$) describe 
the creation (destruction) of phonons with wavevector ${\bf q}$.

In order to provide a gauge-invariant formulation of carrier-phonon 
scattering, let us consider as basis states the 
(gauge-dependent) eigenstates
of the free-carrier Hamiltonian: 
\begin{equation}\label{eigen}
{\bf H}_c \phi_\alpha({\bf r}) = \epsilon_\alpha \phi_\alpha({\bf r}).
\end{equation}
Here, the quantum numbers $\alpha$ ---and therefore the corresponding 
eigenfunctions 
$\phi_\alpha({\bf r}) \equiv \langle {\bf r} \vert \alpha \rangle$ 
and energies $\epsilon_\alpha$--- 
are functions of the transformation parameter $\eta$ 
and in general (i.e., for $\eta \ne 0$) are also functions of time. 
Contrary to the conventional time-dependent perturbation theory, we thus
propose a kinetic 
description based on a time-dependent quantum-mechanical representation. 
In particular, for $\eta = 0$ (scalar-potential gauge) we recover the 
well-known Wannier-Stark ladder \cite{Wannier}: 
$\epsilon_\alpha = \epsilon_n = \epsilon_0 + n \Delta\epsilon$ with
$\Delta\epsilon = eFd$, $d$ denoting the crystal periodicity along the 
field direction; In contrast, for $\eta = 1$ we deal with the Houston or 
accelerated Bloch states \cite{Houston}: 
$\epsilon_\alpha = \epsilon_{{\bf k}(t)}$, 
where ${\bf k}(t) = {\bf k}_0 + \dot {\bf k} t$ is the instantaneous carrier 
wavevector, $\dot {\bf k} = eF/\hbar$ being its field-induced time 
variation.

Generally speaking, we shall denote with ${\cal U}^{\eta,\overline{\eta}}$ the unitary 
transformation connecting the eigenstates $\vert \alpha \rangle$ in 
different gauges: 
\begin{equation}\label{calU}
\vert \alpha(\eta) \rangle = {\cal U}^{\eta,\overline{\eta}} 
\vert \alpha(\overline{\eta}) \rangle \ .
\end{equation}
Given such basis states $\{\vert\alpha \rangle\}$, 
most of the physical quantities we are interested in ---e.g., carrier drift 
velocity and mean kinetic energy--- are properly described 
by the single-particle 
density matrix \cite{DMT}
\begin{equation}\label{DM}
\rho_{\alpha\alpha'} = 
\left\langle 
a^{\dagger}_{\alpha'} a^{ }_\alpha 
\right\rangle,
\end{equation}
where $a^{\dagger}_\alpha$ ($a^{ }_\alpha$) denote creation (destruction)
operators for a carrier in state $\alpha$.
This is defined as the average value of two creation and destruction 
operators:
its diagonal elements 
$f_\alpha = \rho_{\alpha\alpha}$
correspond to the 
usual distribution functions 
of the semiclassical Boltzmann theory 
while the off-diagonal terms ($\alpha\ne\alpha'$) describe 
the degree of quantum-mechanical phase coherence 
between states 
$\alpha$ and $\alpha'$.
It is easy to show that the density matrix (\ref{DM}) will gauge transform
according to:
\begin{equation}\label{rhoeta}
\rho^{\eta}_{\alpha_1\alpha_2} = 
\sum_{\overline{\alpha}_3\overline{\alpha}_4} 
U_{\overline{\alpha}_1\overline{\alpha}_3} 
\rho^{\overline{\eta}}_{\overline{\alpha}_3\overline{\alpha}_4} 
U_{\overline{\alpha}_4\overline{\alpha}_2} \ ,
\end{equation}
where 
$U_{\overline{\alpha}\,\overline{\alpha}'} = \langle \alpha \vert 
\overline{\alpha}' \rangle = 
\langle \overline{\alpha} \vert {\cal U}^{\overline{\eta},\eta} \vert 
\overline{\alpha}' \rangle$
are the matrix elements of ${\cal U}^{\overline{\eta},\eta}$ 
in the $\overline{\eta}$ representation. 
Here, the compact notation 
$\overline{\alpha} \equiv \alpha(\overline{\eta})$ has been introduced. 

For a time-dependent basis set $\{\vert \alpha \rangle\}$,
the Heisenberg equations of motion for the operators 
$a_\alpha$ are of the form \cite{SL}:
\begin{equation}\label{Heisenberg}
\frac{d}{dt} a_\alpha = 
\frac{d}{dt} a_\alpha\Bigl|_{\bf H} +
\frac{d}{dt} a_\alpha\Bigl|_\phi\ .
\end{equation}
Compared to the standard equations of motion, 
the possible time variation of our basis 
states $\phi_\alpha$ gives rise to an additional term; the latter has been 
usually neglected, giving rise to the apparent discrepancies between 
scalar- and vector-potential gauges mentioned in the introductory part of the 
paper (see below). 

By combining Eqs.~(\ref{DM}) and (\ref{Heisenberg}) and considering the 
explicit form of the total Hamiltonian (\ref{H}), we get the following 
equation of motion for $\rho$:
\begin{equation}\label{eom1}
\frac{d}{dt} \rho_{\alpha\alpha'} = 
-i\omega_{\alpha\alpha'} \rho_{\alpha\alpha'} +
\frac{d}{dt} \rho_{\alpha\alpha'}\Bigl|_{{\bf H}_{cp}} +
\frac{d}{dt} \rho_{\alpha\alpha'}\Bigl|_\phi
\end{equation}
with $\omega_{\alpha\alpha'} = 
\left(\epsilon_{\alpha} - \epsilon_{\alpha'}\right)/\hbar$.
The first, Liouville-like, term is due to the single-particle Hamiltonian 
${\bf H}_\circ$ 
while the last one is again due to the possible time 
variation of the basis states $\alpha$. 
The carrier-phonon contribution involves two-body as well as 
various phonon-assisted density matrices, e.g., 
$s_{\alpha\alpha',{\bf q}} = 
\langle a^{\dagger}_{\alpha} b^{ }_{{\bf q}} a^{ }_{\alpha'} \rangle$ 
\cite{DMT}.
These quantities describe many-particle correlations between carriers and 
phonons.
Equation~(\ref{eom1}) is thus the starting point of an infinite hierarchy
involving higher-order density matrices. 
To obtain a solution ---i.e., a closed set of equations--- this hierarchy
has to be truncated at some level. 
As clearly discussed in \cite{DMT}, in order to properly describe 
carrier-phonon scattering, the time evolution of the phonon-assisted 
density matrix $s_{\alpha\alpha',{\bf q}}$ should be explicitely 
considered; its equation of motion has again the form of Eq.~(\ref{eom1}), 
i.e., 
\begin{equation}\label{eom2}
\frac{d}{dt} s_{\alpha\alpha',{\bf q}} = 
-i\Omega^+_{\alpha\alpha',{\bf q}} s_{\alpha\alpha',{\bf q}} +
y^{cp}_{\alpha\alpha',{\bf q}} +
\frac{d}{dt} s_{\alpha\alpha',{\bf q}}\Bigl|_\phi
\end{equation}
with 
$\Omega^\pm_{\alpha\alpha',{\bf q}} =-\omega_{\alpha\alpha'} \pm \omega_{\bf q}$, 
$\omega_{\bf q}$ being the phonon dispersion.
By treating the carrier-phonon contribution 
$y^{cp}_{\alpha\alpha',{\bf q}}$ 
via the standard mean-field approximation \cite{DMT}, 
Eqs.~(\ref{eom1}) and (\ref{eom2}) constitute a closed set 
of equations for the kinetic variables $\rho$ and $s$; they form the basis 
of the so-called {\it carrier-phonon quantum kinetics} \cite{DMT}.

The semiclassical limit is finally obtained via an ``adiabatic 
elimination'' \cite{DMT} of the
phonon-assisted density matrices $s$. This consists in a formal integration of
Eq.~(\ref{eom2}) on which a Markov limit is performed.
More specifically, 
by neglecting the $\phi$-term in Eq.~(\ref{eom2}), 
i.e., the contribution due to the time 
variation of the basis states $\alpha$, the final result is:
\begin{equation}\label{formint}
s_{\alpha\alpha',{\bf q}}(t) = 
{\cal D}(\Omega_{\alpha\alpha',{\bf q}})
y^{cp}_{\alpha\alpha',{\bf q}}(t) 
\end{equation}
with
\begin{equation}\label{D}
{\cal D}(\Omega_{\alpha\alpha',{\bf q}})
= \int_0^\infty dt e^{-i\int_0^t \Omega_{\alpha\alpha',{\bf q}}(t') dt'} \ .
\end{equation}
By inserting the above formal solution for $s$ into the carrier-phonon 
contribution of Eq.~(\ref{eom1}) 
we finally get a closed equation of motion for the 
single-particle density matrix $\rho$.
In the linear regime, i.e., $\left|\rho_{\alpha\alpha'}\right| \ll 1$, 
the carrier-phonon contribution to the dynamics is of the form:
\begin{equation}\label{eom3}
\frac{d}{dt} \rho_{\alpha\alpha'}\Bigl|_{{\bf H}_{cp}}= 
\sum_{\beta \beta'} \left(
\tilde{\Gamma}^{{\rm in}}_{\alpha\alpha',\beta\beta'} \rho_{\beta\beta'} -
\tilde{\Gamma}^{{\rm out}}_{\alpha\alpha',\beta\beta'} \rho_{\beta\beta'} 
\right) 
+ \mbox{c.c.}
\end{equation}
where
\begin{mathletters}
\label{Gammatilde}
\begin{equation}
\tilde{\Gamma}^{{\rm in}}_{\alpha\alpha',\beta\beta'} = \pi 
\sum_{\pm,{\bf q}}
\sum_{\alpha''}
\delta_{\alpha''\beta'}
g_{\alpha\alpha'',{\bf q}}^* 
g_{\alpha'\beta,{\bf q}}
{\cal N}^\pm_{{\bf q}} 
{\cal D}(\Omega^\mp_{\alpha'\beta,{\bf q}})
\end{equation}
and
\begin{equation}
\tilde{\Gamma}^{{\rm out}}_{\alpha\alpha',\beta\beta'} = \pi 
\sum_{\pm, {\bf q}}
\sum_{\alpha''}
\delta_{\alpha'\beta}
g_{\alpha\alpha'',{\bf q}}^* 
g^{ }_{\beta'\alpha'',{\bf q}} 
{\cal N}^\pm_{{\bf q}} 
{\cal D}(\Omega^\pm_{\beta'\alpha'',{\bf q}})
\end{equation}
\end{mathletters}
are generalized in- and out-scattering rates \cite{DMT}.
Here, the $\pm$ sign refers to phonon emission and absorption, 
respectively, 
${\cal N}^\pm_{\bf q} = N_{\bf q} +{1 \over 2} \pm {1 \over 2}$ denote 
the corresponding phonon occupation factors,
and $g_{\alpha\beta,{\bf q}}$ are the matrix elements of the 
carrier-phonon Hamiltonian (\ref{H_cp}). 

Equation (\ref{eom3}) is the desired quantum-mechanical generalization of the 
well-known Boltzmann transport equation \cite{MC}; indeed, by neglecting all 
non-diagonal terms of the single-particle density matrix
($\rho_{\alpha\alpha'} = f_\alpha \delta_{\alpha\alpha'}$), the latter is 
easily recovered:
\begin{equation}\label{BTE}
\frac{d}{dt} f_{\alpha}\Big|_{H_{cp}} =
\sum_{\beta} \left(
\Gamma_{\alpha\beta} f_{\beta} -
\Gamma_{\beta\alpha} f_{\alpha}
\right) \ .
\end{equation}
Here, as usual, the scattering rates for in- and out-scattering processes 
coincide; they correspond to twice the diagonal parts ($\alpha\beta = 
\alpha'\beta'$)
of the scattering operators
$\tilde{\Gamma}^{\rm in}$ and 
$\tilde{\Gamma}^{\rm out}$ in Eq.~(\ref{Gammatilde}):
\begin{equation}\label{FGR}
\Gamma_{\alpha\beta} = \Gamma_{\beta\alpha} = 2\pi
\sum_{\pm,{\bf q}}
\left|g_{\alpha\beta,{\bf q}}\right|^2
{\cal N}^\pm_{{\bf q}} 
\Re\left[{\cal D}(\Omega^\pm_{\alpha\beta,{\bf q}})\right] \ .
\end{equation}
The above semiclassical rates exhibit the well-known structure of the Fermi's 
golden rule; they describe the scattering probability for a phonon-induced 
transition between states $\alpha$ and $\beta$.
Their quantum-mechanical ---or non-diagonal--- generalization is then given 
by the scattering matrices (\ref{Gammatilde}), which describe the effect on the 
time evolution of the density-matrix element $\rho_{\alpha\alpha'}$ due to the 
generic element $\rho_{\beta\beta'}$.

The generalized carrier-phonon scattering rates in (\ref{Gammatilde}) 
---as well as 
their semiclassical counterparts in (\ref{FGR})---
involve the 
${\cal D}$ function in (\ref{D}).
For the case of a time-independent basis set, i.e., $\eta = 0$ 
(Wannier-Stark states), the detuning
frequency $\Omega$ is also time-independent and the real part of the 
function ${\cal D}$ in (\ref{D}) gives the well-known 
energy-conserving Dirac delta 
function of Fermi's golden rule; in contrast, for the case of a 
time-dependent basis, i.e., $\eta = 1$ (accelerated Bloch states), the 
detuning is time-dependent, leading to a function ${\cal D}$ with a 
Fresnel-like shape \cite{Brunetti}.
This is exactly the ICFE previously introduced \cite{BF}: 
due to the field-induced variation of the carrier 
wavevector ${\bf k}$, the energy difference between initial and final 
states 
($\epsilon_{{\bf k}(t)} - \epsilon_{{\bf k}(t)\pm{\bf q}}$)
changes in time giving rise to broad resonances in the carrier-phonon 
scattering process. Such energy-nonconserving scenario has no counterpart 
in the Wannier-Stark picture \cite{CB}. 
This clearly shows that the generalized scattering rates in 
(\ref{Gammatilde}) are
not gauge invariant.

Aim of this paper is to show that (i) the derivation recalled so far 
is only valid within the Wannier-Stark picture ($\eta = 0$) and (ii)
the ICFE previously described is simply an artifact due 
to the approximation scheme usually considered.
Indeed, as anticipated, the crucial point is
the neglect of the $\phi$-terms, i.e., of the possible time 
variation of our basis states $\alpha$. 
More specifically,
a proper inclusion of these terms leads to a modified version of 
Eq.~(\ref{eom2}):
\begin{equation}\label{eom4}
\frac{d}{dt} s_{\alpha\alpha',{\bf q}} = 
i
\sum_{\beta\beta'} \tilde{\Omega}^+_{\alpha\alpha',\beta\beta',{\bf q}} 
s_{\beta\beta',{\bf q}} +
y^{cp}_{\alpha\alpha',{\bf q}} 
\end{equation}
with 
\begin{equation}\label{Omegatilde}
\hbar \tilde{\Omega}^\pm_{\alpha\alpha',\beta\beta',{\bf q}}
= {\cal E}_{\alpha\beta} \delta_{\alpha'\beta'} - 
{\cal E}_{\alpha'\beta'} \delta_{\alpha\beta} \pm 
\hbar\omega_{\bf q} \delta_{\alpha\beta} \delta_{\alpha'\beta'} \ ,
\end{equation}
${\cal E}_{\alpha\alpha'}$ being the matrix elements of the single-particle
Hamiltonian (\ref{H_c}) for $\eta = 0$, i.e., written in the 
scalar-potential gauge.
It follows that for a generic time-dependent basis, Eq.~(\ref{eom4}) 
has a non-diagonal structure,i.e., it does not allow a simple 
exponential solution. This implies that for $\eta \ne 0$ the Markov limit 
is not straightforward. Indeed, the rigorous procedure is: (i) to perform a 
unitary transformation which diagonalizes the superoperator 
$\tilde{\Omega}$ in (\ref{Omegatilde}), and (ii) to perform the exponential
formal integration described above.
Since ${\cal E}_{\alpha\alpha'}$ are the matrix elements of ${\bf H}_c$ for
$\eta = 0$ (scalar-potential gauge), the unitary transformation that 
diagonalizes $\tilde{\Omega}$ is just ${\cal U}^{0,\eta}$, 
i.e., the transformation
connecting  the generic gauge
$\eta$ to the scalar-potential basis ($\eta = 0$).
We stress that the new diagonal elements coincide with the eigenvalues of 
$\tilde{\Omega}$ which, in turn, correspond to the {\it time-independent 
detuning functions 
$\Omega_{\alpha\alpha',{\bf q}}$ in the Wannier-Stark gauge}.
This clearly shows that the Markov limit used to derive the generalized 
Boltzmann equation in (\ref{eom3}) is only well-defined in the Wannier-Stark
picture, for which the various $\phi$--terms vanish and the detuning 
functions
$\Omega$ are time-independent.
This does not violate the gauge-invariant nature of our formulation. 
Indeed, given the generalized Boltzmann equation (\ref{eom3}) written in the 
scalar-potential picture, the latter can be written in any generic gauge 
$\eta$ by applying the unitary transformation ${\cal U}$ introduced in 
(\ref{calU}).
More specifically, 
let us introduce the single-particle density-matrix operator 
\begin{equation}\label{DMO}
\rho = \sum_{\alpha\alpha'} 
\vert \alpha \rangle \rho_{\alpha\alpha'} \langle \alpha' \vert \ ,
\end{equation} 
which is by definition gauge invariant, i.e., $\eta$-in\-de\-pen\-dent. 
This can 
be easily checked by combining Eqs.~(\ref{calU}) and (\ref{rhoeta}).
This suggests to write the generalized Boltzmann equation 
(\ref{eom3}) in an operatorial form as:
\begin{equation}\label{eom3bis}
\frac{d}{dt} \rho\Bigl|_{{\bf H}_{cp}}= \left(
\tilde{\Gamma}^{{\rm in}} \rho -
\tilde{\Gamma}^{{\rm out}} \rho 
\right) 
+ \mbox{h.c.} \ ,
\end{equation}
where 
\begin{equation}\label{Gammatildeo}
\tilde{\Gamma}^{\rm in/out} = \sum_{\alpha\alpha',\beta\beta'}
\vert \alpha \rangle \vert \beta \rangle
\tilde{\Gamma}^{\rm in/out}_{\alpha\alpha',\beta\beta'}
\langle \alpha' \vert \langle \beta' \vert
\end{equation}
are in- and out-scattering superoperators. 
Our aim is to propose a gauge-invariant formulation of the problem. 
This requires the superoperators in (\ref{Gammatildeo}) to be 
$\eta$-independent as well.
The analysis presented so far has shown that the scattering matrices 
$\tilde{\Gamma}^{\rm in/out}_{\alpha\alpha',\beta\beta'}$ are only well 
defined in the Wannier-Stark picture ($\eta = 0$).
This allows us to extend their definitions to the generic $\eta$ 
representation according to:
\begin{equation}\label{Gammatildeeta}
\tilde{\Gamma}^{\eta}_{\alpha_1\alpha_2,\beta_1\beta_2} = 
U_{\overline{\alpha}_1\overline{\alpha}_3} 
U_{\overline{\beta}_1\overline{\beta}_3} 
\tilde{\Gamma}^{\overline{\eta}= 
0}_{\overline{\alpha}_3\overline{\alpha}_4,
\overline{\beta}_3\overline{\beta}_4} 
U_{\overline{\alpha}_4\overline{\alpha}_2} 
U_{\overline{\beta}_4\overline{\beta}_2} 
\ ,
\end{equation}
where $U_{\overline{\alpha}\,\overline{\alpha}'}$ are the matrix elements of 
the gauge transformation ${\cal U}^{0,\eta}$ in the Wannier-Stark picture 
($\overline{\eta} = 0$).
Here, implicit summation over repeated indices is assumed.

Equation (\ref{Gammatildeeta}) is the gauge-invariant formulation of 
Fermi's golden rule we were looking for. 
Contrary to the conventional approach, in the case of a time-dependent 
basis, e.g., accelerated Bloch states \cite{Houston}, instead of using 
Eq.~(\ref{Gammatilde}) with an {\it ad hoc} energy-nonconserving ${\cal D}$ 
function, the rigorous procedure is to compute the generalized scattering 
rates (\ref{Gammatilde}) 
in the Wannier-Stark picture, and then to apply the gauge transformation 
${\cal U}_{\eta,0}$ according to Eq.~(\ref{Gammatildeeta}).

In summary, we have proposed a gauge-invariant formulation of 
carrier-phonon interaction. It has been shown that ICFE ---as usually 
described and accounted for--- is just an artefact due to the neglect of 
the time variation of the basis set.
As anticipated, 
this may account for the surprisingly good agreement between semiclassical 
and rigorous quantum-transport calculations reported in \cite{Brunetti} and
\cite{Herbst}, as well as for the anomalous carrier heating typical of 
semiclassical ICFE models \cite{GMC}. 

From our analysis we can conclude that the most severe 
approximation of the Boltzmann transport theory is not the Markov limit 
but the semiclassical approximation, i.e., the neglect of non-diagonal 
density-matrix elements. The latter, being intrinsically basis dependent, 
is not compatible with a gauge-invariant 
formulation of the problem.


\begin{references}
\bibitem{gauge-invariance}
The need for a gauge-invariant formulation of the problem was originally 
pointed out by Bertoncini and Jauho in \cite{Bertoncini}.
\bibitem{CB}
In addition to ``energy-nonconserving effects'' due to a time-dependent
basis, we also deal with energy-broadening effects, the socalled
``collisional broadening'' \cite{Bertoncini}. 
Such effects can be
easily included in
our theoretical approach by adding an imaginary contribution to the
self-energy $\Omega$ in (\ref{D}).
\end{references}
\end{document}